\documentclass[aps,notitlepage,nofootinbib]{revtex4-2}

\usepackage{bm,amsmath,amssymb,graphicx}
\usepackage{ amsfonts, bm, epsfig, xcolor, algorithm2e}    

\usepackage[format=plain,labelfont={bf,small},textfont=small,justification=raggedright,singlelinecheck=false]{caption}
\usepackage{subcaption}
\begin{document}

\title{Comment on the \emph{elastica} section in Thorne and Blandford \emph{Modern Classical Physics}, the shape of things, and the aspect ratio of reality}

\author{J. A. Hanna} \email{jhanna@unr.edu}%
\affiliation{ Mechanical Engineering, University of Nevada, Reno, NV 89557-0312, U.S.A. }%

\date{\today}

\begin{abstract}
I point out and diagnose an error in a figure in a textbook on classical physics.  The error helps to illustrate a pitfall encountered when dealing with the shapes of objects, and perhaps also reflects general cultural attitudes in physics.  Another, less interesting, error is noted in passing. 
\end{abstract}

\maketitle

This note addresses Figure 11.10 (Section 11.5, Exercise 11.13) of Thorne and Blandford's \emph{Modern Classical Physics} \cite{ThorneBlandford} (also reproduced in \cite{ThorneBlandfordvol3}), which displays incorrect representations of \emph{elastica} equilibria.
The error can be diagnosed as inappropriate affine transformations of correct curves. 
I feel compelled to record this comment, as these curves are a fundamental conceptual and practical building block in my corner of science, and the textbook has a wide audience, being a panoramic introduction to many subfields of classical physics by two well-decorated astrophysicists.\footnote{\ldots who, one might add, began teaching this material at Caltech around the time the presumptuous author of this little comment was born.  It had been my intention to simply contact the authors of \cite{ThorneBlandford} about this error in late 2017, shortly after publication.  That week, however, came the announcement of Thorne's Nobel Prize, so I figured he'd be a bit busy.  I didn't get back to the task until now.} 
It is my hope that this note in defense of a basic ingredient in structural mechanics be taken {\bf{not as criticism, but as a helpful lesson}} about a common and easy type of mistake. It might also call attention, in some small way, to the disconnect between the physics of everyday life and current emphasis in education and research in physics.

Either of the 17th-century planar curve problems, the catenary or \emph{elastica}, might contend to be the ``hydrogen atom'' or ``harmonic oscillator'' of structural mechanics, in that many models of elastic structures are built upon these blocks. 
The \emph{elastica}, an inextensible curve whose energy is quadratic in curvature, subject to forces and moments at its ends, is one of the simplest one-dimensional continua in classical mechanics.  
Its various configurations have been well documented over the last four hundred years or so \cite{Oldfather33, Fraser1991, Levien2008, Love, Antman05, Mladenov17, oreilly2017, SinghHanna19}.  
Subfigures (c) and (d) of Figure 11.10 of \cite{ThorneBlandford} appear jarringly unphysical to anyone who works with thin elastic structures, or is attuned to the presence of this motif in everyday life, whether in the form of electrical cables, plant tendrils, cloth, or sheets of paper. 
To such eyes, the images appear distorted, an impression confirmed below by generating these incorrect curves by unidirectional scaling of correct curves.

One possible equation for static equilibrium of an \emph{elastica} is identical in form to that of the gravitational pendulum, 
\begin{align}
	B d_s^2 \theta + P \sin\theta = 0 \, , \label{pendulum}
\end{align}
where $B$ is a bending modulus, $P$ is the magnitude of applied end forces, $\theta$ is the angle between these forces and the tangent to the curve (defined so that a straight curve will have $\theta = 0$ in compression and $\theta = \pi$ in tension), and $s$ is arc length along the curve.  
This admits the first integral
\begin{align}
	\tfrac{1}{2} B \left(d_s \theta\right)^2 - P \cos\theta = c \, , \label{firstintegral}
\end{align}
where $c$ is the material force or pseudoforce \cite{oreilly2017, SinghHanna19}. 
Solutions are defined, up to uniform scaling, by the ratio $c/P$ \cite{SinghHanna19}. 
Thorne and Blandford consider only cases where no moments are applied, so that the forces are collinear and applied at inflection points of the resulting curves (subfigure (b) of Figure 11.10, which I do not discuss in this note, has not been truncated consistently with this requirement; as pictured in the book, its ends should be subject to both forces and moments).  Thus, only the subset of solutions known as inflectional \emph{elastica} are considered.  The two examples I discuss here are of similar type; 
in both of these shapes the end forces are pulling horizontally outward. 

\begin{figure}[h]
\vspace{0.2in}
    \centering
    \includegraphics[width=6.5in]{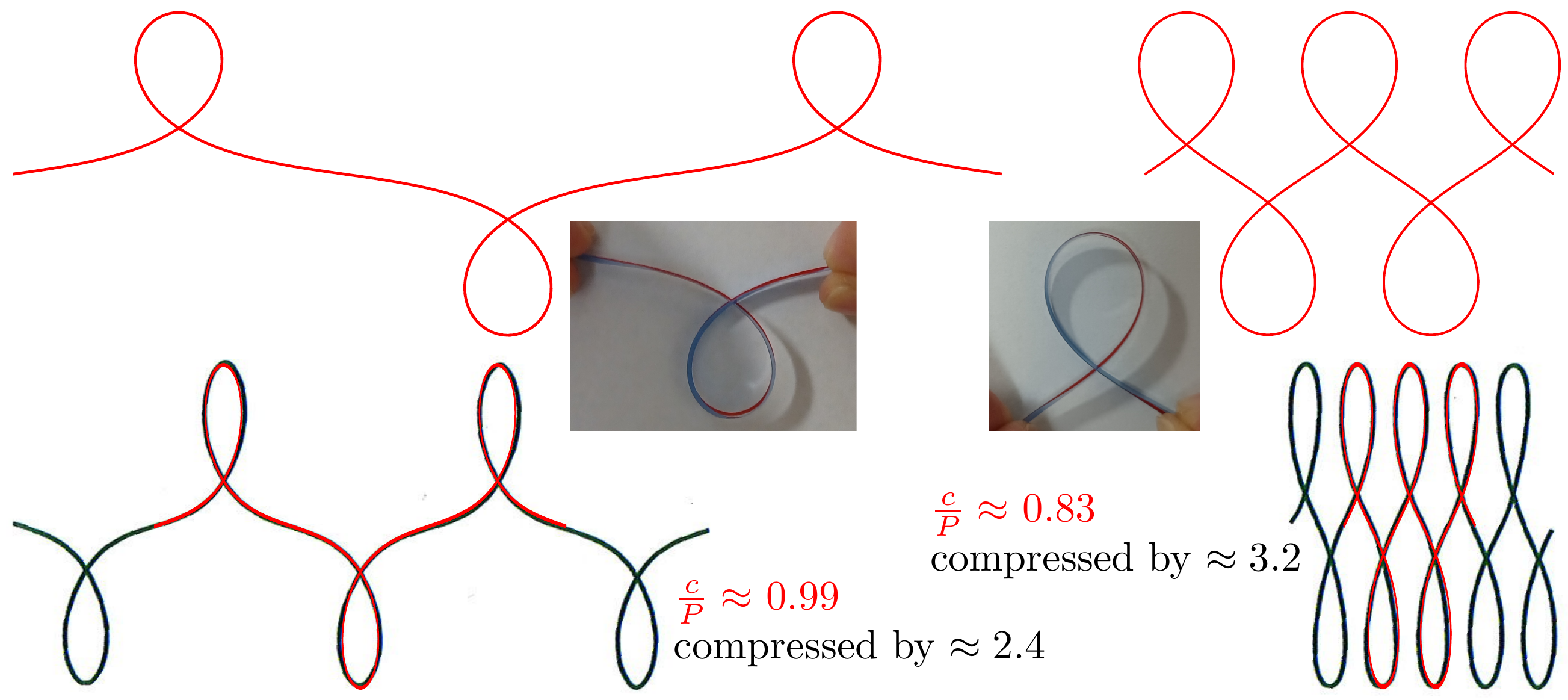}
    \caption{Two \emph{elastica} shapes corresponding to solutions of \eqref{firstintegral} (top, red), overlaid with a thin strip of plastic (insets), and unidirectionally compressed and overlaid on Figure 11.10(c-d) of \cite{ThorneBlandford} (bottom).  Inset photos by A. Dehadrai. 
    %Subfigures from the reference are reproduced with the permission of ....
    }
    \label{compare}
\vspace{0.2in}
\end{figure}

The top of Figure \ref{compare} shows two correct \emph{elastica} configurations as red curves.  That these shapes are correct may be easily verified by comparing them with a bent thin strip of plastic, as in the insets.\footnote{
Those surprised by the utility of such a tabletop ``experiment'' might appreciate Truesdell's comment that 
``Many prefer to mine nature's darkest and deepest entrails, closed except to the dearest experimental apparatus or voluminous statistics, while leaving the smiling face of earth unheeded'' (from \emph{Experience, Theory, and Experiment} (1955), as reprinted in \cite{Truesdell84}). 
%essay first appeared on pages 3-18 of \emph{Proceedings of the Sixth Hydraulics Conference}, Bulletin 36, %State University of Iowa Studies in Engineering, 1956
One reason why \cite{ThorneBlandford} is a celebrated addition to the physics curriculum is that it might counteract the prevailing underemphasis on human-scale phenomena. 
}
The bottom of the figure shows these curves after unidirectional scaling along the line of force, overlaid on subfigures (c) and (d) of Figure 11.10 from Thorne and Blandford.  
This fit by guided trial and error is pretty good, so I conclude that these erroneous images were likely generated either by plotting the curves with an inappropriate aspect ratio, or subsequently by manipulation of subfigures to create a composite figure. 
%error exists in course notes: http://www.pmaweb.caltech.edu/Courses/ph136/yr2012/1211.2.K.pdf
These shapes are not \emph{elastica}, because the transformation  
$x \rightarrow \lambda x$, $y \rightarrow y$, where $\lambda$ is a constant and $x$ and $y$ are coordinates in directions parallel and perpendicular\footnote{Note an important difference in notation whereby \cite{ThorneBlandford} use $x$ to denote our arc length $s$.} to the applied forces (so that $d_s x = \cos\theta$ and $d_s y = \sin\theta$), 
changes the arc length $s$, the angle $\theta$, and derivatives such as the curvature $d_s \theta$, in a manner that does not preserve equations (\ref{pendulum}-\ref{firstintegral}). 
%The relevant physics is not invariant under affine transformations of the embedding space. 
%An easy way to realize this is that the physics is invariant under rigid rotations, which do not commute with affine transformations. 
In general, we should not expect invariance under affine transformations. In structural mechanics, curvature-dependent energies depend on derivatives of angles, and components of forces resolved onto body frames depend on angles.

I have found this mistake to be common among students.
%an important lesson for students 
It is also to be expected from some experienced physicists, who in their usual realm of study are simply not used to plotting \emph{the shape of things}, 
 rather than data or functions whose representation doesn't depend on the anisotropic scaling of a figure. 
Classical mechanics pertains to the everyday reality of life in $\mathbb{E}^3$, and the important lesson here is that \emph{the aspect ratio of reality is unity.}

%I further speculate that some experienced physicists may be surprised by the ease by which a tabletop ``experiment'' can serve as a quick check on such results. Perhaps this is because
%\begin{quote}
%\textsf{Many prefer to mine nature's darkest and deepest entrails, closed except to the dearest experimental apparatus or voluminous statistics, while leaving the smiling face of earth unheeded.} \\
%\small{--- C. Truesdell, \emph{Experience, Theory, and Experiment} (1955), as reprinted in \cite{Truesdell84}. }
%\end{quote}
%%essay first appeared on pages 3-18 of \emph{Proceedings of the Sixth Hydraulics Conference}, Bulletin 36, State University of Iowa Studies in Engineering, 1956
%%  
%%\setlength{\epigraphwidth}{0.62\textwidth}
%%\epigraph{Many prefer to mine nature's darkest and deepest entrails, closed except to the dearest experimental apparatus or voluminous statistics, while leaving the smiling face of earth unheeded.}{C. Truesdell, \emph{Experience, Theory, and Experiment} (1955), as reprinted in \cite{Truesdell84}.}
%Ironically, this is one reason why such a book as \cite{ThorneBlandford} is a celebrated addition to the physics curriculum.

\section*{Acknowledgments}
I thank A. Dehadrai and H. Singh for help.

%\bibliography{refstemp.bib} 
\bibliographystyle{unsrt}

\end{document}